# Emerging edge states on the surface of the epitaxial semimetal CuMnAs thin film

Giang D. Nguyen,[1,2] Krishna Pitike,[3] Peter Wadley,[4] Valentino R Cooper,[3] Mina Yoon[1,6], Tom Berlijn,[1,5*] An-Ping Li[1*]

[1]*Center for Nanophase Materials Sciences, Oak Ridge National Laboratory, Oak Ridge, Tennessee 37831, USA*

[2]*Stewart Blusson Quantum Matter Institute, University of British Columbia, Vancouver, British Columbia V6T 1Z4, Canada*

[3]*Materials Science and Technology Division, Oak Ridge National Laboratory, Oak Ridge, Tennessee 37831, USA*

[4]*School of Physics and Astronomy, University of Nottingham, University Park, Nottingham NG7 2RD, United Kingdom*

[5]*Computational Sciences and Engineering Division, Oak Ridge National Laboratory, Oak Ridge, Tennessee 37831, USA*

[6]*Department of Physics and Astronomy, University of Tennessee, Knoxville, TN 37996, United States*

*\* Corresponding authors: An-Ping Li (apli@ornl.gov) and Tom Berlijn (berlijnt@ornl.gov)

**Abstract**

Epitaxial thin films of CuMnAs have recently attracted attention due to their potential to host relativistic antiferromagnetic spintronics and exotic topological physics. Here, we report on the structural and electronic properties of a tetragonal CuMnAs thin film studied using scanning tunneling microscopy (STM) and density functional theory (DFT). STM reveals a surface terminated by As atoms, with the expected semi-metallic behavior. An unexpected zigzag step edge surface reconstruction is observed with emerging electronic states below the Fermi energy. DFT calculations indicate that the step edge reconstruction can be attributed to an As deficiency that results in changes in the density of states of the remaining As atoms at the step edge. This understanding of the surface structure and step edges on the CuMnAs thin film will enable in-depth studies of its topological properties and magnetism.

Antiferromagnetic (AFM) tetragonal CuMnAs thin films possess many potential advantages in spintronics applications.[1-6] The AFM spin states in CuMnAs can be efficiently and reversibly controlled with an applied electrical current using current-induced spin orbit torques.[2] On the fundamental side, CuMnAs is predicted to be an exotic magnetic Dirac semimetal; where the Dirac crossing in the band structure can be manipulated using electrically controlled magnetic order.[7-9] As a result, CuMnAs is a particularly interesting candidate material for studying the relationship between Dirac fermions and magnetism as well as for exploring topological metal-insulator transitions (MIT) driven by a Néel vector.[7] However, little is known about the surface morphology, intrinsic defects, and electronic properties of the CuMnAs, all which plays an important role for its magnetic properties and device performance, and is a prerequisite for the further understanding of its topological physics. In this regard, scanning tunneling microscopy (STM), an essential technique for probing the local electronic and magnetic structure of materials especially at the surface, is highly desirable for revealing direct correlations between surface morphology, microscopic defects, and electronic and magnetic behaviors.[10-13]

Here we report STM studies of a tetragonal CuMnAs thin film grown epitaxially on a GaP(001) substrate.[2] The STM topography reveal As-terminated surface. The local density of states (LDOS) confirms the semi-metallic behavior of CuMnAs. Interestingly, the clean step edge displays a zigzag periodic structure with a doubling of the in-plane lattice constant, which induces localized states below the Fermi level. DFT calculations reproduce the zigzag step structure and show that the presence of As vacancies are responsible for the emerging edge states at the surface step in the CuMnAs film. This work brings to light the atomic and local electronic structure of CuMnAs surfaces and step edges, inspiring future studies on the topological properties of this material using such as angle-resolved photoemission spectroscopy (ARPES)[14, 15] and the local magnetic structure using spin-polarized STM[16].

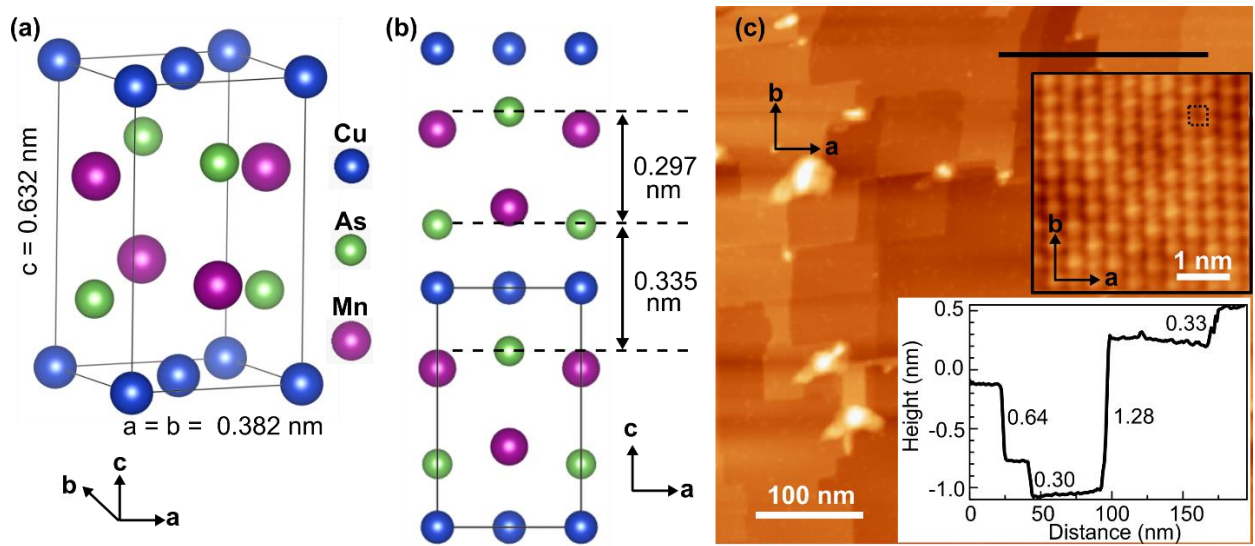

**FIG 1**. (a), (b) Top and side views of the crystal structure of tetragonal CuMnAs. (c) A typical large-scale STM topographic image of the CuMnAs surface after removing the amorphous As capping layer (sample bias $V_s = -2.5$ V, tunneling current $I = 50$ pA). The high-resolution STM image showing its atomic square structure in the inset ($V_s = -1$ V, $I = 50$ pA) (Fourier filtering was applied to enhance the contrast). The lower inset shows the height profile crossing the steps (black line).

A tetragonal CuMnAs thin film was grown epitaxially on GaP(001) substrate, then capped with an As amorphous layer before transferring into our STM system (details in Supplementary Materials). The crystal structure of tetragonal CuMnAs is illustrated in Fig. 1(a), (b). The lattice parameters are c = 0.632 nm and a = b = 0.382 nm. A typical large-scale STM image of CuMnAs after removing the As capping layer is shown in Fig. 1(c). Notably, there is still some residual contamination on the surface as indicated by the bright spots in Fig. 1(c). Nevertheless, the inset in Fig. 1(c) shows an atomically resolved STM image of the CuMnAs surface revealing a square lattice in the **a** and **b** directions with a unit cell of 0.385 ± 0.005 nm, which is half of the diagonal of the GaP unit cell (GaP lattice constant of 0.545 nm) and corresponding to the As-As distance in the (ab) plane of CuMnAs epitaxially grown on a GaP(001) substrate.[17] Step edges are also observed along the **a** or **b** directions, with small deviations due to the thermal drift of the large-scale STM image acquired at room temperature. The lower inset in Fig. 1(c) shows a line profile across several steps taken along the fast scan direction (black line) to avoid the effect from the thermal drift. As expected, step heights appear with an integer number of unit cell size along **c** direction, e.g., 0.64 nm (one unit cell) or 1.28 nm (double of a unit cell). Additionally, step heights of 0.30 ± 0.01 nm and 0.33 ± 0.01 nm are also observed, which are less than a unit cell size but correspond well to the spacing between adjacent As layers as shown in Fig. 1b. Note, the spacings between Mn-Mn layers are 0.215 nm and 0.417 nm, while the spacing of Cu-Cu layers is 0.62 nm. In addition, STM spectroscopy on these different terraces are the same (to be discussed in below), which suggests that they have the same chemical termination. As a result, we determine the top surface of the CuMnAs thin film to be terminated by As atoms.

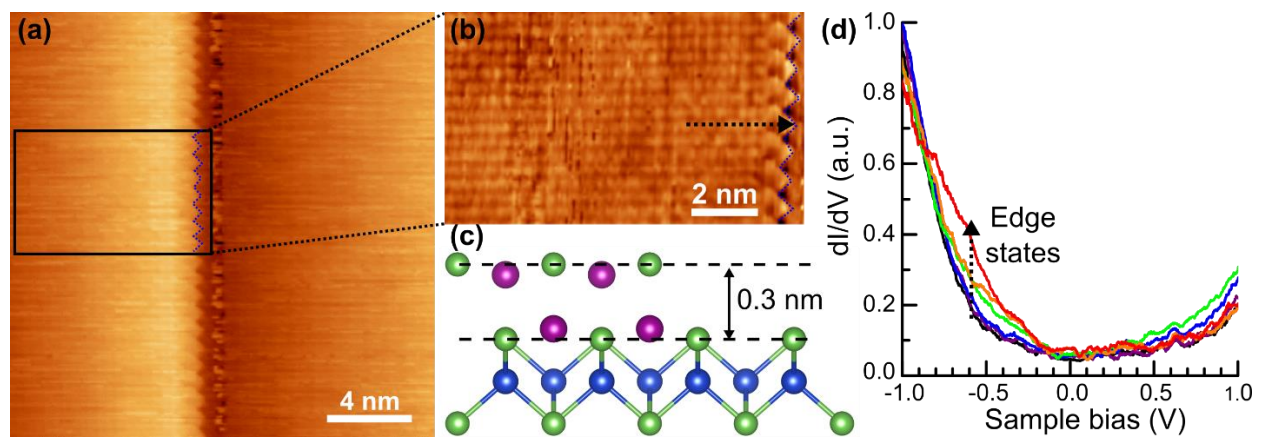

**FIG 2.** (a) STM image across a clean step edge of CuMnAs ($V_s = -1.0$ V, $I = 50$ pA). (b) Atomically resolved zoomed-in STM image at the step edge in (a) ($V_s = -1.0$ V, $I = 50$ pA). (c) A model structure of the step edge. (d) d$I$/d$V$ spectra taken along the black dashed line in (b) showing emerging states below the Fermi level near the step edge ($V_s = -1$ V, $I = 50$ pA, $V_{ac} = 30$ mV, $f = 1000$ Hz). Each curve was averaged from 100 d$I$/d$V$ spectra acquired consecutively to reduce the noise.

Figure 2(a) shows an STM image with a clean step edge (height of 0.3 nm) on the surface. The atomic resolution of As atoms is visible both on the CuMnAs surface and at step edges as displayed in the zoomed-in image of Fig. 2b. Along the step edge, an interesting 1D zigzag feature (blue dashed line) is observed with a periodicity of $0.77 \pm 0.01$ nm which doubles the As-atomic row distance of $0.385 \pm 0.005$ nm. The trench-like feature on the right in Fig. 2a is an artifact induced by a change in the tip-sample interaction when scanning across the step edge. Based on the observed images, a model structure of this step edge is proposed in Fig. 2(c). The d$I$/d$V$ spectra are measured along the line marked in Fig. 2(b) and shown in Fig. 2d near a step edge (black dashed arrow). The d$I$/d$V$ spectra show finite but strong suppression of the DOS is observed at the Fermi level, consistent with the predicted semi-metallic behavior in the CuMnAs samples.[7] Upon approaching the step edge, there is a clear enhancement of the DOS below the Fermi level down to −0.8 eV indicating emergence of new states at step edges.

To better understand the atomic structure near the step edge and emerging edge states we compare the experimental results with DFT calculations. In particular, we simulated an atomic model of the step edge as illustrated in Fig. 2(c). Further details can be found in the Supplementary Materials section. First to reproduce previous literature[18], we examined the thermodynamic stability of their bulk counterparts by computing the corresponding formation energies employing a 4×4×2 CuMnAs supercell with a single Mn or As vacancy. Consistent with Ref.[18] we take the chemical potentials of Mn/As corresponding to Mn/As rich conditions (see Supplementary Materials). The formation energies of As and Mn vacancies in bulk are presented in Table 1. Similar to Ref.[18], we find that the formation energies of As and Mn vacancies in bulk are positive and negative, respectively. This implies that in the bulk the formation of a Mn vacancy in Mn rich conditions is energetically favorable, whereas As vacancies in As rich conditions are not likely to form.

**TABLE 1.** Calculated formation energies for isolated defects in bulk CuMnAs and periodic defects along the step edge. Formation energies for unrelaxed atomic positions are given in parenthesis.

| Defect | Bulk As/Mn rich conditions | | Step Edge T=550 K & P=$10^{-16}$ atm | |
|---|---|---|---|---|
|  | Isolated As vacancy | Isolated Mn vacancy | 1x2 periodic As vacancy | 1x2 periodic Mn vacancy |
| Formation energy [eV] | +1.22 | -0.10 | -0.08 | +1.47 |

Next, we consider two types of edge defects that may give rise to the 1×2 zigzag reconstruction observed in STM: A periodic As vacancy on the upper terrace and a Mn vacancy in the lower MnAs layer (see Figs. S3(b, c)). The CuMnAs sample has been annealed in an ultra-high vacuum (UHV) system before the STM measurements in order to remove the As capping layer. To mimic these conditions, we take the chemical potential of As and Mn to correspond a pressure P = $10^{-16}$ atm (~$10^{-13}$ torr) and a temperature T=550 K (see Supplementary Materials). Interestingly, we find that in this case the energetic favorability of the Mn and As vacancies at a step edge are reversed (Table 1). Along the step edge, Mn vacancies are energetically unfavorable while As vacancies have negative formation energies. Therefore, we conclude that periodic As vacancies are the more likely explanation for the experimentally observed 1×2 reconstruction along the step edge.

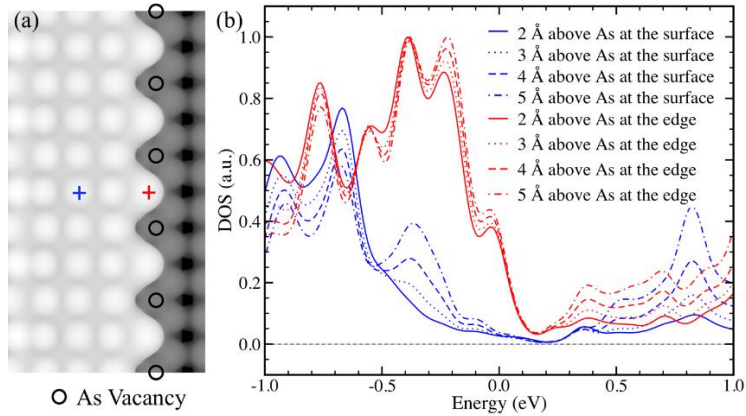

**FIG 3.** (a) STM simulations at −1 V sample bias of a step edge with periodic As vacancies. (b) DOS calculated at 2 to 5 Å above the slab in vacuum for As near the zigzag edge (red cross) and As on the surface (blue cross).

Using the energetically favorable As vacancy step edge model, the STM image of the surface is simulated from our DFT charge density profiles. Figure 3a presents STM simulations for the step edge with periodic As vacancies obtained with the P4VASP software,[19] using an isosurface value of 141.2 Å$^{-3}$. The images show a zig-zag edge structure along the edge of As vacancies which resemble well the STM image shown in Figs. 2(a, b). Furthermore, we plot the DOS calculated in vacuum, 2 to 5 Å above As at the zigzag step edge and As on the surface of the terrace (see Fig. 3(b)). Irrespective of the height, we find a significant increase in the DOS below the Fermi level for As at the step edge, which corresponds very well with our STS data in Fig. 2(d).

Moreover, the contributions to the edge state at –0.5 eV are estimated through analyzing the spin polarized *projected*-DOS (PDOS), shown in Fig. S4. Specifically, we present the PDOS for As and Mn ions near the step edge in the upper MnAs layer. We find that the edge state is predominantly comprised of As-$p$ and Mn-$d$ states, present near the step edge. Interestingly, while As-$p$ states in the bulk and inside the slab have small magnetic moment, ~0.013 µB, As-$p$ states at the surface are significantly spin polarized, with a net magnetic moment of ~0.128 µB. The highly asymmetric nature of the position of the As-ion – arising not only from the step edge but also from the adjacent As-vacancy – could be responsible for the 10-fold increase in the value of local magnetic moment for As ion at step edge compared with the bulk As. Given that CuMnAs can be a Dirac semimetal,[7] these step edge states may have a topologically nontrivial nature.[20] However, further studies are needed to evaluate this behavior.

In summary, we report a detailed study on a tetragonal CuMnAs thin film grown on a GaP substrate using STM and ab initio calculations. The surface is found to be As-terminated. The clean surface has electronic density of states resembling those of a semimetal. The step edges exhibit a zigzag 1×2 reconstruction with edge states below the Fermi level. DFT calculations capture the observed emergent edge states and indicate that the zigzag reconstruction is due to the presence of As vacancies on the step edge.

**Supplementary Material**

See Supplementary Material for methods, and additional figures of atomic model used for DFT calculations and spin polarized PDOS near the step edge.

**Acknowledgement**

This research was conducted at the Center for Nanophase Materials Sciences, which is a DOE Office of Science User Facility. The theory effort (T.B., K. P., V. R. C.) was supported by the U.S. Department of Energy, Office of Science, Basic Energy Sciences, Materials Sciences and Engineering Division. We acknowledge computational resources provided by the National Energy Research Scientific Computing Center (NERSC), which is supported by the Office of Science of the U.S. Department of Energy under Contract No. DE-AC02-05CH11231.

**References**

1. P. Wadley, S. Reimers, M. J. Grzybowski, C. Andrews, M. Wang, J. S. Chauhan, B. L. Gallagher, R. P. Campion, K. W. Edmonds and S. S. Dhesi, Nat. Nanotech. **13**, 362-365 (2018).
2. P. Wadley, B. Howells, J. Železný, C. Andrews, V. Hills, R. P. Campion, V. Novák, K. Olejník, F. Maccherozzi and S. Dhesi, Science **351** (6273), 587-590 (2016).
3. K. Olejník, T. Seifert, Z. Kašpar, V. Novák, P. Wadley, R. P. Campion, M. Baumgartner, P. Gambardella, P. Němec, J. Wunderlich, J. Sinova, P. Kužel, M. Müller, T. Kampfrath and T. Jungwirth, Science Advances **4** (3), eaar3566 (2018).
4. T. Jungwirth, J. Sinova, A. Manchon, X. Marti, J. Wunderlich and C. Felser, Nature Physics **14** (3), 200 (2018).
5. T. Jungwirth, X. Marti, P. Wadley and J. Wunderlich, Nature nanotechnology **11** (3), 231 (2016).
6. M. Grzybowski, P. Wadley, K. Edmonds, R. Beardsley, V. Hills, R. Campion, B. Gallagher, J. S. Chauhan, V. Novak and T. Jungwirth, Physical review letters **118** (5), 057701 (2017).
7. L. Šmejkal, J. Železný, J. Sinova and T. Jungwirth, Physical review letters **118** (10), 106402 (2017).
8. P. Tang, Q. Zhou, G. Xu and S.-C. Zhang, Nature Physics **12** (12), 1100 (2016).
9. L. Šmejkal, Y. Mokrousov, B. Yan and A. H. MacDonald, Nature Physics **14** (3), 242-251 (2018).
10. G. D. Nguyen, J. Lee, T. Berlijn, Q. Zou, S. M. Hus, J. Park, Z. Gai, C. Lee and A.-P. Li, Physical Review B **97** (1), 014425 (2018).
11. G. D. Nguyen, L. Liang, Q. Zou, M. Fu, A. D. Oyedele, B. G. Sumpter, Z. Liu, Z. Gai, K. Xiao and A.-P. Li, Physical Review Letters **121** (8), 086101 (2018).
12. C. Friesen, H. Osterhage, J. Friedlein, A. Schlenhoff, R. Wiesendanger and S. Krause, Science **363** (6431), 1065 (2019).

13. S. M. Hus and A.-P. Li, Progress in Surface Science **92** (3), 176-201 (2017).
14. Z. K. Liu, B. Zhou, Y. Zhang, Z. J. Wang, H. M. Weng, D. Prabhakaran, S. K. Mo, Z. X. Shen, Z. Fang, X. Dai, Z. Hussain and Y. L. Chen, Science **343** (6173), 864 (2014).
15. N. P. Armitage, E. J. Mele and A. Vishwanath, Reviews of Modern Physics **90** (1), 015001 (2018).
16. R. Wiesendanger, Reviews of Modern Physics **81** (4), 1495-1550 (2009).
17. P. Wadley, V. Novák, R. Campion, C. Rinaldi, X. Martí, H. Reichlová, J. Železný, J. Gazquez, M. Roldan and M. Varela, Nature communications **4**, 2322 (2013).
18. F. Máca, J. Kudrnovský, V. Drchal, K. Carva, P. Baláž and I. Turek, Physical Review B **96** (9), 094406 (2017).
19. http://www.p4vasp.at.
20. L. Peng, Y. Yuan, G. Li, X. Yang, J.-J. Xian, C.-J. Yi, Y.-G. Shi and Y.-S. Fu, Nature Communications **8** (1), 659 (2017).